\documentclass[iop]{emulateapj}

\def\kms{\ifmmode km\thinspace s^{-1}\else km\thinspace$s^{-1}$\fi}     
\def\deg{\ifmmode^\circ\else$^\circ$\fi}  
\def\arcs{\ifmmode {'' }\else $'' $\fi}  
\def\arcm{\ifmmode {' }\else $' $\fi}    

\def\msun{M$_\odot$}

\def\hi{\ion{H}{1}}
\def\lapp{\ifmmode {_<\atop{^\sim}} \else {$_<\atop{^\sim}$}\fi} 
\def\gapp{\ifmmode {_>\atop{^\sim}} \else {$_>\atop{^\sim}$}\fi} 

\shorttitle{Metallicity of the Leo Ring}
\shortauthors{Rosenberg et al.}

\begin{document}

\title{Unravelling the Mysteries of the Leo Ring: An Absorption Line Study of an
Unusual Gas Cloud\altaffilmark{*}}

\author{J. L. Rosenberg\altaffilmark{1} and Karl Haislmaier\altaffilmark{1,2}}
\affil{School of Physics, Astronomy, and Computational Science, George Mason University,
    Fairfax, VA 22030, USA}
\affil{Department of Astronomy, University of Massachusetts, Amherst, MA 01003, USA}
\email{jrosenb4@gmu.edu}

\author{M. L. Giroux}
\affil{Department of Physics and Astronomy, East Tennessee State University, Johnson City, TN, 37614, USA}

\author{B. A. Keeney}
\affil{Center for Astrophysics and Space Astronomy, Department of Astrophysical and Planetary Sciences,
University of Colorado, 389 UCB, Boulder, CO 80309, USA}

\and

\author{S. E. Schneider}
\affil{Department of Astronomy, University of Massachusetts, Amherst, MA 01003, USA}

\altaffiltext{*}{Based on observations obtained with the NASA/ESA
{\it Hubble Space Telescope}, which is operated by the Association
of Universities for Research in Astronomy, Inc., under NASA
contract NAS5-26555.  These observations are associated with
program GO12198.01-A.}

\begin{abstract}
Since the 1980's discovery of the large ($2 \times 10^9$ \msun) intergalactic
cloud known as the Leo Ring, this object has been the center of a lively debate
about its origin. Determining the origin of this object is still important as
we develop a deeper understanding of the accretion and feedback processes that
shape galaxy evolution. We present HST/$COS$ observations of three sightlines
near the Ring, two of which penetrate the high column density neutral hydrogen
gas visible in 21
cm observations of the object. These observations provide the first direct
measurement of the metallicity of the gas in the Ring, an important clue to its
origins. Our best estimate of the metallicity of the ring is $\sim$10\%
Z$_{\odot}$, higher than expected for primordial gas but lower than expected
from an interaction. We discuss possible modifications to the interaction and
primordial gas scenarios that would be consistent with this metallicity
measurement. 
\end{abstract}


\section{INTRODUCTION}

\par Over twenty-five years ago a giant ring of neutral hydrogen gas (diameter
of $\sim$200 kpc and \hi\ mass of 2 $\times 10^9$ M$_{\odot}$) was found in the
Leo region \citep{schneider1983, schneider1985, schneider1986, schneider1989a,
schneider1989b}. At the time, the ring was subject to speculation about its
origins -- whether it is the remnant of a tidal interaction or is primordial gas
left over from the formation of the group of galaxies in which it sits. 

The 21 cm observations provide gas velocities so the original observations of
the ring indicated a velocity gradient around the ring that implied that the
ring is rotating around a kinematic center that is $15 \pm 7$ \kms\ from the
luminosity weighted center of M105. The highest velocities of the gas are
$\sim$1100 \kms\ in the
southwest and the lowest velocities of $\sim$750 \kms\ in the northeast of the
ring (\citealp{schneider1985,alfalfa40},Osterloo et al. private communication).
The gas velocities translate to an orbital period of $4 \times 10^9$ years.


\par Since that time, the ring has been the subject of additional studies aimed
at providing clues to its origins. Multiwavelength observations of the ring have
shown strong limits on the stellar emission within the ring. Early  V and K-band
observations set the surface brightness limits, at the 3 highest density peaks
identified in \citet{skrutskie1984}, as 28.0 and 22.8 mag arcsec$^{-2}$
respectively. This lack of significant stellar emission in the ring indicates
that if it is a remnant of the disruption of a galaxy, the region of the galaxy
that was disrupted had to be very gas-rich with a low optical surface
brightness.

More recent $r$-band (from the KPNO 4-meter) and FUV and NUV (from
GALEX) observations identify faint overdensities of sources in the vicinity of
the highest \hi\ column density region. Population synthesis modeling of these
optical/UV overdensities are claimed to indicate metallicities of Z$_{\odot}$/50
\citep{thilker2009}. Such low metallicities are consistent with a
primordial origin for the ring but not with a tidal origin.
However, in addition to the measurement of metallicity from population
synthesis, there is a tentative detection of dust emission at 8$\mu$m in
the densest region of the ring \citep{bot2009}. The detection, if confirmed,
implies a gas-to-dust ratio between 1/20 and a few times solar. This level of
dust is inconsistent with the low metallcity derived by \citet{thilker2009}
and is much more consistent with a tidal origin for the gas. 

\par In addition to NGC 3384, M105, M96, and NGC 3351, the bright galaxies in the
vicinity of the ring, there are 6 lower luminosity galaxies that have been
identified \citep{stierwalt2009}. For two (AGC 201970 and AGC
202027) of these 6 galaxies, the optical velocities place them at a redshift
consistent with that of the ring. A third galaxy (AGC 205505) has an optical
velocity just above that of the ring but still within the group. The remaining 3
galaxies do not have optical redshifts so their velocities, relative to that of
the ring, are not known. \citet{stierwalt2009} suggest that these galaxies may
have formed in situ as tidal dwarf galaxies. These galaxies do not help resolve
the question of the origin of the gas since their metallicity is unknown. The
existence of tidal dwarf galaxies associated with the Ring is consistent with
both scenarios of the feature's origin and would aid in our understanding
of the feature if their metallicity was known. 

\par While the studies of \citet{thilker2009} and \citet{silchenko2003} have
argued that the gas is primordial, the connection between the gas in the Ring
and the galaxies NGC 3384 and M 96 (as observed in the 21 cm map, see Figure
\ref{fig:qsopos}) is often used as evidence of a tidal origin. \citet{rood1985}
propose a head-on or Spitzer-Baade collision between
NGC 3384 and M105 while \citet{michel-dansac2010} also argue for a head-on
collision but consider NGC 3384 and M96 to be the more likely candidates. 
Alternatively, \citet{bot2009} argue that their mid-infrared detection of dust
associated with the densest part of the ring is consistent with the gas left
over from the stripping of a low surface brightness galaxy as has been modeled
by \citet{bekki2005}.

The global kinematics have provided additional constraints on the origin of the
Leo Ring. Observations of the Ring show a smooth gradient of velocity from
$\sim$1100 \kms\ in the southwest down to $\sim$650 \kms\ in the northeast.
\citet{schneider1985} have modeled the velocities as a rotating ring of gas with
a kinematic center that is 15 $\pm$ 7 kpc from the luminosity weighted centroid
of the M105/NGC 3384 system and an orbital period of $4 \times 10^9$ years,
longer than the crossing time of the system. On the scale of the individual
clouds, the gas exhibits a high velocity disperion that probably arises from
multiple blended clumps along the line of sight. In addition, there is evidence
for a more diffuse medium between the clumps \citep{schneider1986}.

\par Evidence has been accumulating on both sides of the debate about the
origins of this unusual object without resolving the conflict. In that context,
we present Cosmic Origins Spectrograph (COS) observations of three QSOs
that lie behind the ring. We use these observations to study the metallicity and
ionization properties of the ring. \S 2 discusses the COS observations of these
3 sightlines, \S 3 describes the data reduction that was done, \S 4 presents
ionization modeling of the sightlines, and \S5 provides a discussion of the
results. Throughout the paper we use a distance of 10.4 Mpc for the ring, the
average distance to the five large
galaxies in the region determined from well-established distance indicators
\citep{harris2007}.

\section{OBSERVATIONS}

\begin{figure}[h]
\epsscale{1.2}
\plotone{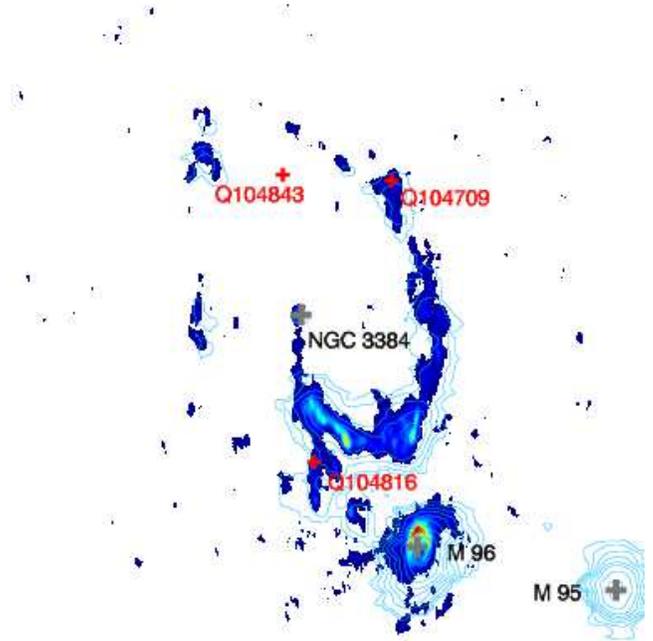}
\caption{\hi\ 21cm observations of the Leo Ring. The color scale shows the \hi\
column density from the Westerbork map (Osterloo et al. priv. comm.). Contours
are from the Arecibo telescope as part of the ALFALFA survey
\citep{stierwalt2009,alfalfa40}. The contour levels are at 5, 15, 30, 50, 120,
300, and 400 $\times 10^{20}$ cm$^{-2}$. The positions of M96 and M95 are
labeled in the map. Note that M95 is outside of the area covered by the
Westerbork map. The red circles indicate the positions of the three QSOs behind
the Leo Ring.}
\label{fig:qsopos}
\end{figure}

We have obtained HST/$COS$ spectra at the positions of three QSOs that lie behind
the Leo Ring. Table \ref{tab:tab-1} lists the QSOs and their positions are shown
overlaid on the Westerbork (Osterloo et al. priv. comm.) and Arecibo
\citep{alfalfa40} \hi\ 21 cm maps of the
Leo Ring in Figure \ref{fig:qsopos}. We observe SDSS J104816.25+120734.7
(Q104816 hereafter) and J104709.83+130454.6 (Q104710 hereafter) with the G130M
grating and observe SDSS J104843.49+130605.9 (Q104843 hereafter) with both the
G130M and G160M gratings. The details of the observations including the exposure
times in each grating are given in Table \ref{tab:tab-1}. 

\par Q104816 probes the dense gas ($log$ N$_{HI} \sim 19.5$ cm$^{-2}$ as
measured in the Westerbork emission map) at the Southern edge of
the ring a bit south of the densest region region where young stars have been
detected by GALEX \citep{thilker2009} and dust may have been detected by Spitzer
\citep{bot2009}. Q104709 probes a slightly lower column density (log N$_{HI} \sim
19.1$ cm$^{-2}$) region than Q104816 in the Northwest region of the Ring.
Q104843 is not behind the detected
\hi\ emission contours, but it is just below the ring on the Northern edge (see
Figure \ref{fig:qsopos}). 

\subsection{Absorption Line Measurements using HST/$COS$ Spectra}

The \textsc{corrtag} files were extracted from the archive and one-dimensional
spectra were extracted using \textsc{CalCOS v2.19.4}. Previous versions of
\textsc{CalCOS} calculate errors for the flux values using
the Gaussian approximation for Poisson errors ($\sigma_i \approx \sqrt{N}$,
where $N_i$ is the gross counts at pixel $i$). This assumption systematically
underpredicts the errors for small values of $N_i$, and since our targets are
quite faint ($\gtrsim90$\% of the pixels in every exposure have $N_i \leq 5$) it
is important that we treat the errors in this regime appropriately.  Starting
with version 2.19.4, \textsc{CalCOS} uses the formalism of \citet{gehrels86},
which was specifically designed to be valid for all values of $N_i \geq 0$, to
determine the Poisson errors for the flux values.

After the extraction, cross correlation and coaddition of the one-dimensional
spectra were then performed with custom routines developed by the COS GTO team
specifically for
COS FUV data\footnote{IDL routines for coadding COS data are available at
\url{http://casa.colorado.edu/$\sim$danforth/science/cos/costools.html}.} as
described in \citet{danforth10} and \citet{keeney12}. After the creation of the
one-dimensional spectra, continua are fit to the coadded spectra using a
semi-automated routine described in \citet{keeney13}.  These continuum fits are
used to normalize the data for easier absorption line fitting.  

Measurements of the \ion{O}{1}, \ion{N}{1}, and the Ly$\alpha$ line require
additional processing of the spectra to mitigate the effects of terrestrial
airglow. The modest recession velocity of the Leo Ring
($cz\approx800$--$900$ \kms) toward our QSO sight lines causes the \ion{O}{1}
$\lambda1302.2$~\AA, \ion{N}{1} $\lambda1199.5, 1200.2$, and $1200.7$~\AA\ and
Ly$\alpha$ $\lambda1215$~\AA\ absorption features to
fall in regions of the spectrum that suffer from terrestrial airglow emission. 
To identify the times when the terrestrial airglow was significant we used the
\textsc{CORRTAG} files returned from the archive. Orbital night was defined
to be times when the zenith distance from the sun was $>95\degr$. When the
separation from the sun was larger than this value, the contribution from
\ion{O}{1} $\lambda1304.9$~\AA\ airglow emission was, emipirically, found to be
minimized.  The \textsc{corrtag} files
for each exposure were then split into separate files for orbital day and
orbital night and one-dimensional spectra were extracted using \textsc{CalCOS
v2.19.4}.  The resulting coadded spectra for each sight line incorporate the
full (orbital day + night) exposure time for the vast majority of wavelengths
and night-only data in regions affected by terrestrial airglow emission.  This
procedure does not completely remove \ion{H}{1} Ly$\alpha$ airglow emission from
the coadded spectrum but it virtually eliminates the \ion{O}{1} and \ion{N}{1}
airglow emission, at the expense of larger error values in these regions.
The spectra created from these split data were then used to estimate
limits on the \hi\ Ly$\alpha$ absorption for the three sightlines.
We were also able to make a measurement of N(\ion{N}{1}) along the
Q104816 sightline, but neither a measurement nor a limit on N(\ion{O}{1}) was
possible. We were unable to make measurements or obtain meaningful limits on
N(\ion{N}{1}) or N(\ion{O}{1})  for the other sightlines.

Voigt profile fits to the absorption lines of
interest were performed using custom IDL routines
that convolve the idealized Voigt profile with the COS on-orbit line spread
function \citep{kriss11} to properly account for instrumental resolution
effects.  All atomic data needed to generate Voigt profiles with a given Doppler
parameter and column density were taken from \citet{morton03}. The results of
the line fitting are presented in Tables \ref{tab:tab-2}, \ref{tab:tab-3}, and
\ref{tab:tab-4} for the sightlines Q104816, Q104710, and Q104843 respectively.
From the fit to each of the lines, column density, recession velocity, and
doppler parameter measurements along with their uncertainties are returned. We
use single component fits to the data for these lines. Fit parameters SL, b,
$\log$ N, velocity and the associated errors are only listed once for each
atomic species since all of the lines of that species were fit simultaneously.
Details of the spectral
fitting and evaluation of the uncertainties can be found in \citet{keeney13}.
Tables \ref{tab:tab-2}, \ref{tab:tab-3}, and \ref{tab:tab-4} also include a
measurement of the significance level,
$SL$, of each line following the formalism outlined in \citet{keeney12}. 

\section{IONIZATION MODELING OF SIGHTLINES}

We use the measurements of metal-line absorption along the three QSO sightlines
to constrain the amount of ionized gas and the metallicity associated with the
Leo Ring gas. We adopt solar values for metal abundances (on a logarithmic
scale) from \citet{asplund2009} of [H]=$12.00$, [C]=8.43, [N]=7.83, [O]=8.69, and
[Si]=7.51.

Because we only observe select ionization states for the elements that we are
examining, we must make ionization corrections to calculate the total column
density of a given element from the measured column density of a single
ionization state. As an example, the ionization corrections for Si are defined as: 

\begin{equation}
\mbox{[Si/H]} = \mbox{[\ion{S}{2}/\ion{H}{1}]} + (correction) 
\label{eq:ioncorr}
\end{equation}

where logarithmic quantities are implied. Such corrections are uncertain even if
the gas is primarily photoionized because they depend on the gas density and the
properties of the ionizing radiation field. We also neglect the possibility of
depletion. 

In order to estimate ionization corrections we have constructed a grid of
photoionization models associated with each sightline using the Cloudy
modelling code \citep[version 10.00]{Ferland98}.  We model the absorption as a
single, constant density (one phase), plane parallel slab illuminated by a uniform
incident radiation field.  We consider two spectral shapes for the incident
radiation field based upon the calculations of the metagalactic radiation field
of Haardt \& Madau (as implemented within Cloudy based upon personal
communication from those authors). In the first model, the incident radiation
field includes emission from both QSOs and galaxies.
This model assumes the contribution from galaxies to the
hydrogen ionizing spectrum is approximately equal to that
from QSOs. The incident radiation field in the second model (``QSO-only'') 
assumes a background solely due to QSOs.  This model has a
higher proportion of high energy photons and may represent
a limit in which few ionizing photons escape from galaxies.

The ionization corrections are dependent on the ionization
parameter $U = n_\gamma / n_H$, where
$n_\gamma = \Phi / c$ is the density of photons above
the hydrogen ionization potential, and $n_H$ is the hydrogen
density.  The quantity $\Phi$ is the surface flux of hydrogen
ionizing photons (photons~${\rm cm}^{-2}{\rm s}^{-1}$).
The surface flux $\Phi_{HM}$ of hydrogen ionizing photons assumed
in the first Haardt \& Madau model (galaxies plus QSOs) is $\Phi_{HM} =
3.2\times 10^4 ~{\rm photons}~{\rm cm}^{-2}{\rm sec}^{-1}$.  This value depends
on the level of the metagalactic radiation field
at $z = 0$, and is a quantity that remains uncertain.
The level of the metagalactic radiation field has been
expressed in terms of $\Phi$ as well as in terms of
$\Gamma$, the hydrogen photoionization rate in units of s$^{-1}$, and $J_{-23}$,
the mean intensity in units of
$10^{-23}~{\rm erg}~{\rm cm}^{-2}~{\rm s}^{-1}~{Hz}^{-1}~{\rm ster}^{-1}$,
and constraints are usually expressed in terms of
one of these quantitities.  There is a weak spectral
dependence on the relation between these values,
but approximately, for $\Phi_{HM} = 3.2\times 10^4$,
$\Gamma \approx 7\times 10^{-14}$, and $J_{-23} \approx 3$.

For the cases where we have measurements or limits on higher
ionization species such as \ion{Si}{3}, \ion{Si}{4}, and \ion{C}{4}, we
determine whether the values are consistent with the conditions determined for
the lower ionization species.
Because these more highly ionized species are often associated with a different
phase of the gas we do not apply the results directly but consider whether they
provide evidence for a more highly ionized phase.

If we assume $\Phi$, (or $\Gamma$ or $J_{-23}$)
is known, we can translate the ionization parameter $U$
to an inferred density $n_H$ and, in combination with an observed \hi\ column
density, can translate this into a line-of-sight extent
of the gas.  Assuming that the line-of-sight extent of the
gas is unlikely to greatly exceed the lateral extent, we can constrain the
allowed range of $U$. The success of this exercise depends on how well the level
of the metagalactic background at $z = 0$ is known. Estimates of this
level have been made based upon the integration of the quasar luminosity
function, the line of sight proximity effect, the flux decrement method, and
from limits on H$\alpha$ emission associated with isolated gas clouds.
The results of many of these investigations are summarized in
\citet[see their Figure 6]{adams2011}. In general, these estimates of the
metagalactic background are somewhat lower than the fiducial value that is used
in the Cloudy models, $\Phi/\Phi_{HM} \lapp 0.7$. For a given ionization parameter
$U$, the line of sight extent of the gas scales roughly with
the inverse of $\Phi$.  As a result, we assume that
estimates of the line of sight extent based upon $\Phi_{HM}$
represent conservative lower limits on the extent of the gas; thus the
assumption that the line of sight extent of our model is $<30$~kpc places a
lower limit on the range of likely ionization parameters.

\subsection{Absorption Line Results for SDSSJ 104816.25+120734.7}

\begin{figure}[h]
\plotone{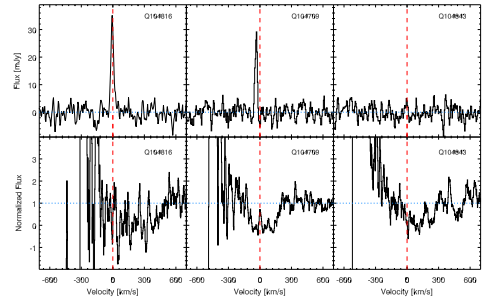}
\caption{The \hi\ 21 cm emission lines as measured from the ALFALFA survey
\citep{alfalfa40} for each sightline are shown in the top panel. The
corresponding Ly$\alpha$ absorption in the HST/$COS$ spectra, smoothed to
$\sim$12 \kms\ are shown in the bottom panels. Because of the Ly$\alpha$ airglow 
emission seen at velocities less than 0 and the Milky Way Ly$\alpha$
absorption, we use the radio \hi\ measurements even though the resolution is much
lower for these data.}
\label{fig:hiplot}
\end{figure}

Q104816 lies behind a radio contour corresponding to $\log N(HI) = 19.5$ on the
\hi\ 21 cm Westerbork map (Osterloo et al. priv. comm.). The angular resolution
of the \hi\ emission observations are significantly lower than the HST FUV
observations -- $1.75\arcm \times 0.65\arcm$  for the radio map in contrast with
the subarcsecond pencilbeam associated with the UV absorption line observations.
Despite the resolution difference, we do not measure the Ly$\alpha$ absorption
directly from the HST/$COS$ spectra that are shown in Figure \ref{fig:hiplot}
along with the \hi\ 21 cm emission profiles from the ALFALFA survey
\citep{alfalfa40} for this and the other 2 sightlines discussed below. We do not
use the Ly$\alpha$ absorption to measure the \hi\ content because it is strongly
blended with absorption from Milky Way Ly$\alpha$ making a measurement from
these data highly unreliable. However, for the Q104816 sightline, the HST/$COS$
data does indicate that the Westerbork value, $\log$ N(\hi) = 19.5 is consistent
with the Ly$\alpha$ absorption profile. 

\begin{figure}[h]
\plotone{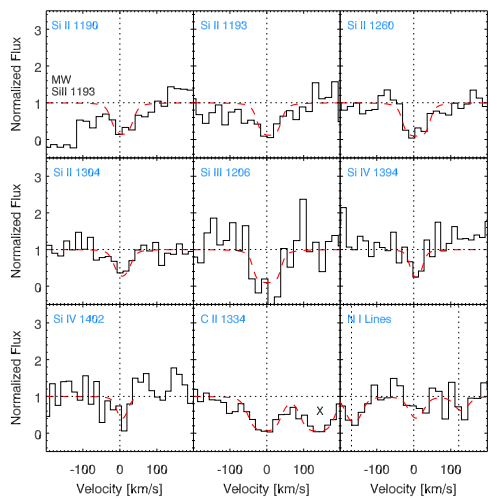}
\caption{The absorption lines identified and measured in the Q104816 spectrum.
The red dashed lines indicate the fits to the lines that were used. The spectra
are centered at a velocity of 917 \kms.}
\label{fig:lines104816}
\end{figure}

Figure \ref{fig:lines104816} shows the Q104816 HST/$COS$ spectra and absorption
line fits for the \ion{Si}{2} $\lambda$1190, 1193, 1260, 1304, \ion{Si}{3} $\lambda
1206$, \ion{Si}{4} $\lambda 1394, 1402$, \ion{C}{2} $\lambda 1334$, and
\ion{N}{1} $\lambda 1199, 1200$ lines on which we base our results. 

The absorption line measurements, line velocities, significance level (SL) of
the line fits, and the integrated EW values are given in Table \ref{tab:tab-2}.
We note that the weighted average of the line velocities is 917 \kms, 5 \kms
larger than the value measured from the Westerbork \hi\ 21 cm emission. This
difference is within the expected uncertainty in the velocity scale for the
HST/$COS$ measurements. Since we do not make use of the line velocities in our
analysis, this difference will not have an effect on our results. The resulting
column densities along this line of sight are $\log$ N(\ion{Si}{2}) = 14.4 $\pm
0.5$ cm$^{-2}$, $\log$ N(\ion{C}{2}) $= 14.7 \pm 0.1$ cm$^{-2}$ and $\log$
N(\ion{N}{1})$= 14.3 \pm 0.4$ cm$^{-2}$.

\subsubsection{A Metallicity Estimate}

We use the measurements of the absorption lines to determine our nominal
measurement of the metallicity along this sightline. We use the \hi\ column
density from the Westerbork 21 cm emission line measurement, $\log$ N(\hi)
$= 19.5$. This value gives column density ratios with respect to \hi\ of
[\ion{Si}{2}/\hi] $= -5.1 \pm 0.5$, [\ion{C}{2}/\hi] $= -4.8 \pm 0.1$, and
[\ion{N}{1}/\hi] $= -5.2 \pm 0.4$. 
If the fractions in these ionization stages matched the fraction of neutral
hydrogen (i.e., no ionization correction is needed), this would imply
logarithmic abundances with respect to solar of $-0.6 \pm 0.5$, $-1.2 \pm 0.1 $,
and $-1.0 \pm 0.4$ for silicon, carbon, and nitrogen, respectively.

\begin{figure}[h]
\plotone{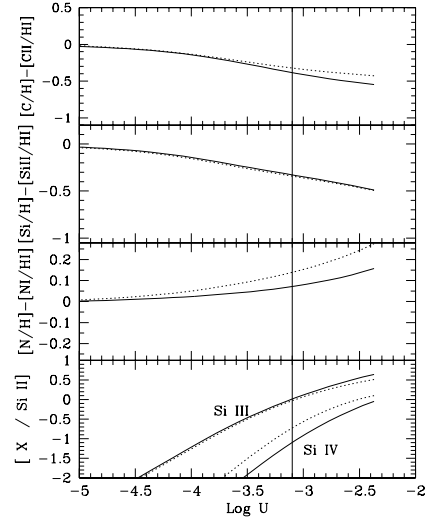}
\caption{The top three panels represent ionization corrections
for C, Si, and N, respectively (see Equation \ref{eq:ioncorr}) versus
$\log U$ where $U$ is the ionization parameter. The solid curves assume the AGN
plus galaxies metagalactic radiation field while the dotted curves assume
an AGN-only shape to the radiation field. The bottom
panel represent model ratios of \ion{Si}{3}/\ion{Si}{2} and \ion{Si}{4}/\ion{Si}{2},
as labeled, versus $\log U$.  The vertical line through all
panels represents the maximum allowed value of $\log U$
assuming a maximum line of site extent of $30$~kpc and
$\Phi_{HM} = 3.2\times 10^4 {\rm photons}~{\rm cm}^{-2}{\rm sec}^{-1}$ (see \S 4).
All models assume $\log N(HI) = 19.5$, appropriate for the
Q104816 sightline.}
\label{fig:104816_ioncor}
\end{figure}

Figure \ref{fig:104816_ioncor} shows the results of a set of photoionization
models designed to determine the ionization corrections required to obtain
elemental abundances $[Si/H]$, $[N/H]$, and $[C/H]$. When the ionization
corrections, as plotted in Figure \ref{fig:104816_ioncor} (and Figure
\ref{fig:104709_ioncor}), are negative as in the cases of silicon and carbon,
the metallicity corrected for ionization will be lower than the measured ratios.
Alternatively, when the ionization correction is positive, as it is for
nitrogen, the metallicity will be higher than the measured ratio. Requiring the
line of sight extent of the gas to be less than $30$~kpc constrains the gas to
have $log U < -3.1$  for $\Phi_{HM} = 3.2 \times 10^4$ phot cm$^{-3}$ sec$^{-1}$
and $log N$(\hi) $= 19.5$. The largest ionization corrections required are
$-0.4$, for [\ion{S}{2}/\hi] and [\ion{C}{2}/\hi] and $0.2$ for
[\ion{N}{1}/\hi] so the metallicities corrected for ionization are $[Si/H] =
\left. -5.5^{+0.9}_{-0.4} \right.$, $[C/H] = \left. -5.2^{+0.5}_{-0.1} \right.$,
and $[N/H] = \left. -5.0^{+0.4}_{-0.6} \right.$. The errorbars on these
measurements conservatively include the ionization corrections since they are
uncertain. Expressing these values in terms of solar abundances, these values
correspond to $[Si/H]_\odot = \left. -1.0^{+0.9}_{-0.4} \right.$,
$[C/H]_\odot = \left. -1.6^{+0.5}_{-0.1} \right.$, and $[N/H]_\odot = \left.
-0.8^{+0.4}_{-0.6} \right.$. 

For comparison, \citet{schneider1986} measured the properties of gas clumps
identified in the ring. They inferred densities for the clumps of -1.8 $< \log
n_{HI} < -0.6$ cm$^{-3}$. However, since these clumps represent \lapp 0.5 of
the integrated flux they also infer a more diffuse medium with $\log n_{HI}$
\lapp -3 cm$^{-3}$. Our values for $U$ and $\Phi$ correspond to $\log n(H) >
-2.9$ cm$^{-3}$ which is approximately $\log n_{HI} > -4.3$ cm$^{-3}$,
consistent with either a dense phase or a more diffuse phase. Our ionization
corrections assume that the gas is associated with the more
diffuse phase, but the denser phase would imply smaller ionization corrections
and thus higher abundances of carbon and silicon (and marginally lower
abundances of nitrogen).

The measured values for the ratios, [\ion{Si}{3}/\ion{Si}{2}] $= 1.6 \pm 1.4$,
and [\ion{Si}{4}/\ion{Si}{2}] $= -0.6 \pm 0.3$, are just barely consistent with
the models at $\log n_H \approx -3$ for an AGN plus stellar ionizing background.

\subsection{Absorption Line Results for SDSSJ 104709.83+130454.6}

Q104709 lies behind a radio contour corresponding to $\log N(HI) = 19.1$ on the
\hi\ 21 cm Westerbork map (Osterloo et al. priv.comm.). A nominal fit to the
Ly$\alpha$ absorption indicates a column density of $\log N(HI) = 18.8 \pm 0.3$.
As with the previous sightline, this region of the spectrum has a very low S/N
due to the damping wings of Galactic Ly$\alpha$ and a smaller exposure time
because daytime data were eliminated to reduce the Ly$\alpha$ airglow. Given the
difficulties with the absorption line measurement, we adopt the Westerbork
measurement of $\log N(HI)$ rather than the measurement from the HST/$COS$
spectrum as we did for Q104816 because it is strongly blended with absorption
from Milky Way Ly$\alpha$. The spectrum is shown in Figure \ref{fig:hiplot}
along with the ALFALFA emission profile at this position. 

\begin{figure}[h]
\plotone{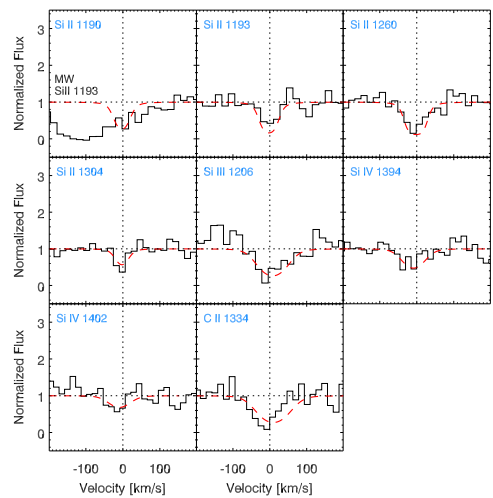}
\caption{The metal line absorbers identified and measured in the HST/$COS$
spectrum of Q104709. The red dashed lines show the fits to the lines that
were used. The velocity is centered at 855 \kms.}
\label{fig:lines104709}
\end{figure}

Figure \ref{fig:lines104709} shows the HST/$COS$ spectrum and absorption line
fits for the \ion{Si}{2} $\lambda 1190, 1193,$ and $1260$, \ion{Si}{3} $\lambda
1260, 1304$, \ion{Si}{4} $\lambda 1394$ and $1402$, and \ion{C}{2} $\lambda
1334$ lines on which we base our results. 

The absorption line measurements, line velocities, significance level (SL) of
the line fits, and the integrated EW values are given in Table \ref{tab:tab-3}.
We note that the weighted average of the line velocities is 857 \kms, 38 \kms
larger than the value measured from the Westerbork \hi\ 21 cm emission. This
difference is larger than the expected uncertainty in the velocity scale for the
HST/$COS$ measurements, 15 \kms, possibly a combination of the velocity scale
uncertainty and the low S/N of the data. 

\subsubsection{A Metallicity Estimate}

We use the measurements of the absorption lines to determine our nominal
measurement of the metallicity for Q104709 as we did for Q104816.  The resulting
column densities along this line of sight are $\log$ N(\ion{Si}{2})$ =
14.0 \pm 0.2$ cm$^{-2}$ and $\log$ N(\ion{C}{2})$ = 14.5 \pm 0.1$ cm$^{-2}$. For
$\log$ N(\hi) = 19.1, the column density ratios with respect to \hi\ are
[\ion{Si}{2}/\hi] $= -5.1 \pm 0.2$, [\ion{C}{2}/\hi] $= -4.6 \pm 0.1$. 

\begin{figure}
\plotone{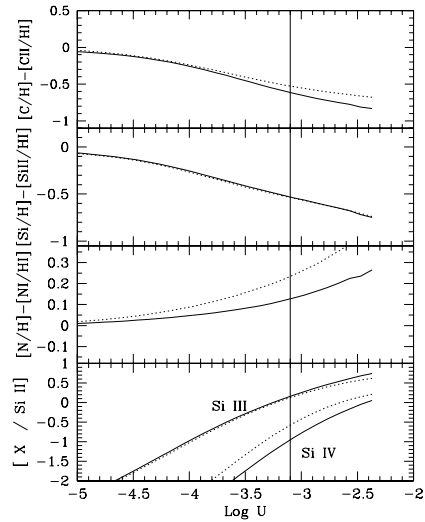}
\caption{The top three panels represent ionization corrections
for C, Si, and N, respectively (see Equation 2) versus
$\log U$ where $U$ is the ionization parameter.  The bottom
panel shows model ratios of \ion{Si}{3}/\ion{Si}{2} and \ion{Si}{4}/\ion{Si}{2}, as
labeled, versus $\log U$.  The vertical line through all panels represents the
maximum allowed value of $\log U$ assuming a maximum line of site extent of
$30$~kpc and $\Phi_{HM} = 3.2\times 10^4 {\rm photons}~{\rm cm}^{-2}{\rm
sec}^{-1}$ (see \S 4). All models assume $\log N(HI) = 19.1$, appropriate for
the Q104709 sightline.}
\label{fig:104709_ioncor}
\end{figure}

Figure \ref{fig:104709_ioncor} shows the photoionization models used to compute
the ionization corrections that were applied to obtain elemental
abundances $[Si/H]$ and $[C/H]$ assuming $\log$ N(\hi)$ = 19.1$, Requiring the
line of sight extent of the gas to be less than $30$~kpc constrains the gas to
have $log U < -3.1$  for $\Phi_{HM} = 3.2 \times 10^4$ phot cm$^{-3}$ sec$^{-1}$
and $log N$(\hi) $= 19.1$.

The largest ionization corrections consistent with our line of sight constraint
are of -0.5 and -0.6 for silicon and carbon respectively. This would imply
logarithmic abundances with respect to solar of [Si/H] $= \left.
-1.1^{+0.7}_{-0.2} \right.$ and [C/H] $= \left. -1.0^{+0.7}_{-0.1} \right.$
While the \ion{Si}{3} absorption line is not highly significant, we accept
the estimate of $\log$ N \ion{Si}{3} $= 13.4 \pm 0.2$ cm$^{-2}$ and associate it
with the same phase that contains \ion{C}{2} and \ion{Si}{2}. In this case, the
bottom panel of Figure \ref{fig:104709_ioncor} shows that the ionization
parameter would be less than $log U \approx -3.5$. A small value for $log
U$ would reduce the ionization required and would increase the abundance
estimates for silicon and carbon.

While the \ion{Si}{3} and \ion{Si}{4} absorption lines are not highly
significant, if their relative strengths are taken at face value they provide
strong evidence for the presence of a multiphase medium. As the bottom panel of
Figure \ref{fig:104709_ioncor} indicates, a very low gas density (high $U$) is
required for the \ion{Si}{3} and \ion{Si}{4} column densities to be comparable
in a single phase medium. Alternatively, the relative \ion{Si}{4} column density
could be elevated by assuming that it is in a higher temperature phase than that
associated with pure photoionization equilibrium. This warm diffuse phase may
have a negligible contribution to the \hi\ column density but may contain the
vast majority of the \ion{Si}{4} and possibly \ion{Si}{3} column density along
the sightline.

The presence of a higher ionization phase along the sightline should not
strongly impact our metallicity estimates associated with the low
ionization lines. To the extent that the \ion{Si}{3} column density is
associated with the higher ionization phase rather than the low ionization
phase, it may indicate a smaller ionization correction. This smaller ionization
correction, when applied to the low ionization phase, would then imply silicon
and carbon abundances approaching the high end of our estimated range. 

\subsection{SDSSJ 104843.49+130605.9}

No radio emission has been detected at the position of Q104843 implying an upper
limit on the average column density of $log$ N(\hi) $< 18.73$ cm$^{-2}$,
although regions within the radio beam could have higher column density if the
medium is clumpy. While a definitive measurement of N(\hi) from the Ly$\alpha$
absorption is not possible, we have compared absorption profiles with a range of
N(\hi) and b-values with the data and have concluded that values of N(\hi)
substantially above (or below) $log$ N(\hi) $= 18.73$ cm$^{-2}$ are not
compatible with the HST/$COS$ data.

\begin{figure}[h]
\plotone{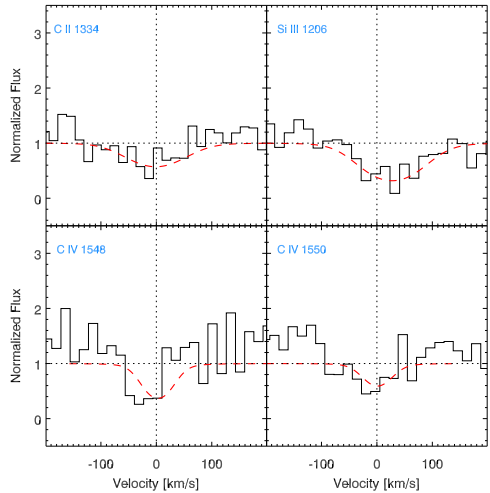}
\caption{The absorption lines identified and measured in the Q104843 spectrum.
The red dashed lines indicate the fits to the lines that were used. The velocity
is centered at 844 \kms.}
\label{fig:lines104843}
\end{figure}

Figure \ref{fig:lines104843} shows the \ion{C}{2} $\lambda
1334$, \ion{Si}{3} $\lambda 1206$, and \ion{C}{4} $\lambda 1548$ and 1550 lines
that are detected along this sightline and the spectral fits upon which we base
our tabulated velocities, doppler parameters and column densities along with their
formal uncertainties. The summary of our results is given in Table
\ref{tab:tab-4}. Given the low formal SL's associated with these metal lines,
it is difficult to say anything quantitative about the metallicity
of the gas along this sightline. The detection of \ion{C}{4} along this
sightline provides further indication of a warm diffuse gas component within the
Leo Ring and extended beyond the boundaries of the cold neutral component
observed in \hi.

\subsection{Summary}

We have measured metal-line absorption along three QSO sightlines though the
Leo Ring. For the two sightlines that show \hi\ 21 cm emission, we make use of
those detections to compute N$_{HI}$. This is done despite the beam size
mismatch between the two observations because of the uncertainty in the
Lyman $\alpha$ absorption measurements (see \S 3.1 and \S 3.2 for details).
 
Using the photoionization models discussed in the text we derive the metallicity
along the highest column density sightline, Q104816 to be between 2 and 10\%
solar. Along the sightline Q104709 we find a nominal metallicity of 10-16\%
solar and a lower limit of 2\% solar. The third sightline, Q104843, shows metal
line absorption, but a quantitative determination of the metallicity can not be
made because of the lack of a reliable \hi\ detection.

The measurements of high ionization species in Q104709 and Q104843 indicate the
existance of a higher ionization phase of the gas than what is probed by the
lower ionization species. While the higher ionization species in Q104816 are
consistent with the lower ions, it may be that there is a higher ionization
phase, it is just not significant enough to affect the line ratios.

The uncertainties in our measurements of metallicity are largely from the lack
of a Lyman$\alpha$ measurement of the \hi\ column density of these sightlines and
the low signal-to-noise of the data.

\section{DISCUSSION}

This study of the metallicity in the Leo Ring was designed to constrain the
origin of the gas. Using the absorption lines identified in the QSO spectra, we
find a metallicity of $\sim$10\% Z$_{\odot}$, marginally higher than the 2\%
Z$_{\odot}$ found by \citet{thilker2009} but consistent with the measurements of
\citet{bot2009} which, if correct, indicate that there is dust in the ring. We
explore the implications of this metallicity with regard to three scenarios for
the formation of the Ring: (1) a tidal interaction between M96 and NGC 3384, (2)
the disruption of a low surface brightness galaxy by the group, and (3)
primordial intergalactic gas.

The discussion of these three scenarios is based on our measurements of the
metallicity along these sightlines which are summarized in Table
\ref{tab:metals}. These values have been measured as well as our data will
allow, but there is some uncertainty beyond the measured errors. This
uncertainty is because the QSO sightlines are not well matched spatially by the
21 cm beam used to measure the \hi\ column density. The values measured from the
Westerbork maps are consistent with the Ly$\alpha$ absorption, to the limits
of the QSO data, but the uncertainties are large. If N(\hi) is larger than we
have assumed, the ratios of [\ion{Si}{2}/\hi] and [\ion{C}{2}/\hi] would be
smaller so uncorrected for ionization the abundances would be lower. However,
for a given ionization parameter, the ionization correction for a larger \hi\
column density has a smaller absolute value which, in the cases of silicon and
carbon, where the ionization corrections are negative, would at least partially
offest the decrease in the abundance. However, nitrogen has a positive
ionization correction so the errors would tend to add, but the ionization
corrections are generally small so the error is dominated by the uncertainty in
the \hi\ column density. 

The metallicity in gaseous tidal tails is difficult to determine directly so it
is often measured in associated tidal dwarf galaxies. These tidal dwarf galaxies
are found to have high metallicities (often near solar or even super-solar;
e.g., \citealp{tf2012}, \citealp{demello2012}) for their luminosities -- they do
not follow the luminosity-metallicity relation (e.g., \citealp{duc2014}). This
deviation from the relation is probably because material from which they formed
is drawn from the inner disk of
the galaxy. \citet{michel-dansac2010} have modeled the Leo Ring as material
removed when M 96 plunged though the center of NGC 3384 $\sim$1.2 Gyr ago. NGC
3384 has a metallicity of 1.1 times solar at 1.25 kpc from the center of the
galaxy \citep{sb2007} and M96 is a very metal-rich galaxy with a metallicity of
$\sim$3.2 Z$_{\odot}$ 3.3 kpc from the center of the galaxy \citep{oey1993}. The
enrichment found in tidal dwarf galaxies would suggest that if the Leo Ring was
formed as a tidal tail resulting from an interaction between these galaxies a
metallicity near solar, significantly larger than what we have measured here,
would be expected. 

However, the Leo Ring, with its circular morphology, looks more like a
collisional ring (i.e., the result of a head-on collision between NGC 3384 and
M105) than a more generic tidal interaction. \citet{appleton1996} review the
models of collisional rings and show that the rings come from the expansion of
the outer regions of the disk and are thus less affected by metallicity mixing
than other tidal interactions. In a study of 8 collisional rings
\citet{bransford1998} find metallicities ranging from about quarter solar to
solar, closer to but still somewhat higher than found for the Ring. In addition
to this metallicity difference, it is also important to note that the Leo Ring
is significantly larger than any other collisional ring galaxy known (where the
largest are $\lapp 60$ kpc, \citealp{madore2009}) and it lacks the star
formation seen in the models and observations of other collisional rings
(e.g., \citealp{higdon1997, fogarty2011, mapelli2012}). 

As an alternative to looking at the metallicities of tidal dwarf galaxies, which
may not be an appropriate model if the material was not drawn from the inner
disk of the galaxy, a
lower limit on the expected metallicity due to a tidal interaction can be
found by extrapolating from the metallicity in the inner region of the disk to
the distance of the Ring assuming a metallicity gradient for the galaxy. Since
one would expect the gas to arise at smaller radii than the current location of
the ring and for some mixing to occur, the projected metallicity of the galaxies
at large radii provides a lower limit. Given the central metallicity and
gradient for NGC 3384 \citep{sb2007}, the galaxy would reach $\sim$20\% solar
metallicity at 30\arcmin\ (90 kpc) from the center, the approximate distance of
the Ring. The galaxy would not reach 10\% solar metallicity until $>800$ kpc
(500 kpc if you use the elevated central metallicity in this calculation). For M
96, assuming the most extreme metallicity gradient (-0.25) seen in the galaxies
sampled by \citet{werk2011} (with the exception of the lower slope for NGC 2146)
the metallicity of M 96 would be $\sim$11\% at 90 kpc. So assuming that the
metallicity is correct for the Ring, it is very difficult to match the
metallicity with gas associated from the galaxy without a much steeper
metallicity gradient in the outer regions of these galaxies than seen in the
inner regions of the disk. In most of the galaxies where metallicity has been
studied at large radii flattening of the metallicity gradient is observed
(e.g.,\citealp{werk2011,bresolin2012,bresolin2009,goddard2011}).

Interactions seem to be the most obvious source for the gas in the Leo Ring --
morphologically the gas appears to be connected to an interaction between NGC
3384 and M 96. The tidal scenarios discussed above imply a metallicity that is
higher than what has been calculated from the absorption lines
that we have measured. Even with the uncertainties in our metallicity
measurements, it is difficult to reconcile the low metallicities that we
calculate with what we would expect for tidal material surrounding super-solar
metallicity galaxies. One possibility in reconciling this model with the data
is that there has been significant mixing between tidal material and more
primoridal gas. To acheive a metallicity of 10\%, the enriched
material removed from these systems would have to be a minority component mixed
with low metallicity material from the IGM -- in the extreme case where gas that
is only 20\% solar is removed from the galaxy and mixed with 3\% Z$_{\odot}$ gas
from the IGM, 59\% of the gas would have to come from the IGM to create the 10\%
Z$_{\odot}$ gas in the Ring.  

Another scenario that has been proposed for the Ring is the disruption of a low
surface brightness galaxy. The galaxy has to have been low luminosity and low
surface brightness for there to be no trace of the stars in optical observations
of the Ring. Such a system, because of the luminosity-metallicity relationship
(e.g., \citealp{tremonti2004}),
would be expected to be low metallicity. While the inability to detect stars
from the galaxy that was disrupted is problematic for this scenario, the
metallicity of the gas would be consistent. For comparison, several new
metallicity measurements along the Magellanic Stream give a value of 10\% solar,
consistent with material removed from the dwarf irregular galaxy, the Small
Magellanic Cloud (SMC; \citealp{fox2013}). However, the \hi\ mass of the SMC is
only 4.2 $\times 10^8$ \msun \citep{stanimirovic1999}, significantly smaller
than the \hi\ mass of the Leo Ring. 

The final scenario, initially proposed by \citet{schneider1985} is that the gas
in the Ring is primordial. \citet{silchenko2003} also argued for this scenario
after studying the stellar ages and kinematics of the galaxies in the group.
They looked at the stellar ages and kinematics of the central
regions of NGC 3384, M 96, and M 105 and concluded that all three galaxies have
central disks or rings. They argue that those features are aligned, or in the
case of the ring in M 105, orthogonal to the line of nodes of the Leo Ring. They
use these kinematics and the mean ages of the central stellar populations, which
are all $\sim$3 Gyr and close to the 4 Gyr orbit time for the Ring, to argue
that the Ring is primordial and is still feeding material into the galaxies.
The metallicity result of 2\% solar found by \citet{thilker2009}
would be consistent with this scenario. While our nominal measurement of the
metallicity of the Ring is higher than expected given this scenario, we can not
rule out a metallicity of a few percent. However, our nominal measurement of
metallicity seems to indicate that if the Leo Ring is IGM material, it has been
polluted, at least to some degree by material from the surrounding galaxies.

\citet{lehner2013} and \citet{wotta2014} studied the metallicity of Lyman-limit
systems at
$z < 1$ and find that they have a bimodal distribution with peaks at $\sim$2.5\%
and $\sim$50\% solar metallicity. The higher metallicity material is thought to
trace winds, recycled material, and outflows while the more pristine gas is the
more primordial material. The gas in the Ring falls between the two branches
which may imply a rare mixing of primordial and enriched material. If M 105, the
elliptical in the group, was responsible for the enrichment of the gas through
outflows that transformed this galaxy to its current gas-poor state, we can
look at the metals created in the formation of the galaxy's stellar population
to gain insight into the possible enrichment. Using the
population synthesis models of \citet{bruzualcharlot} and assuming that the
stellar population of this galaxy was formed in a single burst, the amount of
gas returned to the ISM from type II SNe is $2.4 - 2.5 \times 10^{10}$ \msun,
with the range reflecting differences in possible starting metallicities.
Given that the metal yield of SNe averaged over a population of massive stars is
expected to be roughly solar in metallicity \citep{woosley2002}, only $\sim
1.4 \times 10^8$ M$_{\odot}$ of this gas needs to be removed from M105 and mixed
with 3\% Z$_{\odot}$ metallicity material in the IGM to enrich the Ring to the
measured level. We use 3\% as the nominal IGM metallicity based on the low
metallicity branch of the Lyman-limit cloud distribution measured by
\citet{lehner2013}.

Using absorption lines towards three QSOs behind the Leo Ring we have measured
the nominal metallicity of the Ring to be $\sim$ 10\% solar. This metallicity is
lower than expected for tidal material stripped in an interaction between M96
and NGC 3384, the primary
candidates for such an interaction. Alternatively, it would be consistent with
the expected metallicity of a low surface brightness galaxy, but this scenario
suffers from the the lack of identification of an optical component to such a
system and the need to consider an extremely gas-rich interaction candidate. The
nominal metallicity is also higher than expected for primordial
material that has not been polluted by outflows and other recycled material but
consistent with primordial material that has been mixed with more metal-enriched
material. While the uncertainties in our data do not allow us to rule out any of
the proposed scenarios they do indicate that the gas being tidal in origin,
which seems most likely morphologically, is hard to reconcile with the
measurements without at least some significant mixing with low metallicity
material. Nevertheless, the data also seem to indicate that
pollution of the gas by the galaxies through tidal mixing or outflows is likely,
i.e., that the gas is not completely pristine.

\acknowledgments

This work was supported by the Hubble Grant, HST-GO-12198.01-A. We thank Curtis
Struck for his extremely helpful feedback and suggestions for this work. We
appreciate Tom Osterloo for his willingness to share the Westerbork maps of the
Leo Ring and thank Sabrina Stierwalt and the ALFALFA collaboration (PIs Haynes and
Giovanelli) for the Arecibo
maps of the Leo Ring and for useful discussions.Brian Kent and
John Feldmeir also provided insightful discussions that helped shape this
project. 

{\it Facilities:} \facility{HST (COS)} 

\begin{deluxetable*}{cccr}
\tablecaption{COS Observations}
\tablehead{
Target & z & Grating & $t_{exp}$ (s)}
\startdata
SDSSJ104816.25+120734.7 & 0.2909 & G130M &  14384    \\
SDSSJ104709.83+130454.6 & 0.3997 & G130M &  8383    \\
SDSSJ104843.49+130605.9 & 0.2185 & G130M &  11435   \\
SDSSJ104843.49+130605.9 & 0.2185 & G160M &  11389   
\enddata
\label{tab:tab-1}
\end{deluxetable*}

\begin{deluxetable*}{rrrrrrrrrrr}
\tabletypesize{\scriptsize}
\tablecaption{SDSSJ 104816.25+120734.7 Absorption Measurements}
\tablehead{
\colhead{Ion} & $\lambda$ & \colhead{SL} & \colhead{EW} & \colhead{$\sigma_{EW}$} &  \colhead{b} &
\colhead{$\sigma_b$} & \colhead{log N} & \colhead{$\sigma_N$} & \colhead{Vel} &
\colhead{$\sigma_{Vel}$} \\
& [\AA] & & [m\AA] & [m\AA] & [km/s]  & [km/s] & [cm$^{-2}$] & [cm$^{-2}$] & [km/s]  & [km/s]
}
\startdata
\ion{N}{1}  & 1199.55 & 2.8     & 147 & 116 &  19.9    & 21.7    &  14.3   & 0.4     & 924     & 15	    \\
\ion{N}{1}  & 1200.22 & \nodata & 89  & 119 &  \nodata & \nodata & \nodata & \nodata & \nodata & \nodata   \\
\ion{N}{1}  & 1200.71 & \nodata & 59  & 85  &  \nodata & \nodata & \nodata & \nodata & \nodata & \nodata   \\
\ion{Si}{2} & 1190.42 & 7.5     & 255 & 153 & 19.1     & 5.4	 & 14.4    & 0.5     & 923     & 6	    \\
\ion{Si}{2} & 1193.29 & \nodata & 445 & 233 & \nodata  & \nodata & \nodata & \nodata & \nodata & \nodata   \\
\ion{Si}{2} & 1260.42 & \nodata & 319 & 133 & \nodata  & \nodata & \nodata & \nodata & \nodata & \nodata   \\
\ion{Si}{2} & 1304.37 & \nodata & 175 & 115 & \nodata  & \nodata & \nodata & \nodata & \nodata & \nodata   \\
\ion{C}{2}  & 1334.53 & 2.4     & 485 & 55  & 65.7	& 9.2	  & 14.7    & 0.1     & 907	& 7	    \\
\ion{Si}{3} & 1206.50 & 3.6     & 328 & 194 & 20.0	& 0.0	  & 14.3    & 1.8     & 918	& 23	    \\
\ion{Si}{4} & 1393.75 & 2.4     & 113 & 136 & 13.7	& 20.8    & 13.6    & 0.9     & 923	& 13	    \\
\ion{Si}{4} & 1402.77 & \nodata & 164 & 192 & \nodata   & \nodata & \nodata & \nodata & \nodata & \nodata 
\enddata
\label{tab:tab-2} 
\end{deluxetable*}

\begin{deluxetable*}{rrrrrrrrrrr}
\tabletypesize{\scriptsize}
\tablecaption{SDSSJ 104709.83+130454.6 Absorption Measurements}
\tablehead{
\colhead{Ion} & $\lambda$ & \colhead{SL} & \colhead{EW} & \colhead{$\sigma_{EW}$} &  \colhead{b} &
\colhead{$\sigma_b$} & \colhead{log N} & \colhead{$\sigma_N$} & \colhead{Vel} &
\colhead{$\sigma_{Vel}$} \\
& [\AA] & & [m\AA] & [m\AA] & [km/s]  & [km/s] & [cm$^{-2}$] & [cm$^{-2}$] & [km/s]  & [km/s]
}
\startdata
\ion{Si}{2} & 1190.42 & 9.6     & 206 & 102 & 19.1    & 3.4     & 14.0    & 0.2	  & 852     & 4       \\
\ion{Si}{2} & 1193.29 & \nodata & 132 & 65  & \nodata & \nodata & \nodata & \nodata & \nodata & \nodata \\
\ion{Si}{2} & 1260.42 & \nodata & 236 & 76  & \nodata & \nodata & \nodata & \nodata & \nodata & \nodata \\
\ion{Si}{2} & 1304.37 & \nodata & 89  & 42  & \nodata & \nodata & \nodata & \nodata & \nodata & \nodata \\
\ion{C}{2}  & 1334.53 & 3.4     & 257 & 79  & 47.0    & 14.2    & 14.5    & 0.1	  & 864     & 1       \\
\ion{Si}{3} & 1206.50 & 2.5     & 257 & 124 & 42.0    & 23.0    & 13.4    & 0.2	  & 862     & 1       \\
\ion{Si}{4} & 1393.75 & 2.5     & 110 & 59  & 34.9    & 15.6    & 13.5    & 0.2	  & 845     & 1       \\
\ion{Si}{4} & 1402.77 & \nodata & 86  & 55  & \nodata & \nodata & \nodata & \nodata & \nodata & \nodata 
\enddata
\label{tab:tab-3} 
\end{deluxetable*}

\begin{deluxetable*}{rrrrrrrrrrr}
\tabletypesize{\scriptsize}
\tablecaption{SDSSJ 104843.49+130605.9 Absorption Measurements}
\tablehead{
\colhead{Ion} & $\lambda$ & \colhead{SL} & \colhead{EW} & \colhead{$\sigma_{EW}$} &  \colhead{b} &
\colhead{$\sigma_b$} & \colhead{log N} & \colhead{$\sigma_N$} & \colhead{Vel} &
\colhead{$\sigma_{Vel}$} \\
& [\AA] & & [m\AA] & [m\AA] & [km/s]  & [km/s] & [cm$^{-2}$] & [cm$^{-2}$] & [km/s]  & [km/s]
}
\startdata
\ion{C}{2}  & 1334.53 & 1.5     & 170 & 151 & 64.9    & 46.8    & 14.2    & 0.3	  & 842     & 1       \\
\ion{Si}{3} & 1206.50 & 2.3     & 312 & 159 & 64.4    & 32.1    & 13.4    & 0.2	  & 871     & 23      \\
\ion{C}{4}  & 1548.19 & 1.8     & 192 & 157 & 30.7    & 31.9    & 14.0    & 0.4	  & 845     & 1       \\
\ion{C}{4}  & 1550.78 & \nodata & 170 & 263 & \nodata & \nodata & \nodata & \nodata & \nodata & \nodata 
\enddata
\label{tab:tab-4} 
\end{deluxetable*}

\begin{deluxetable*}{cccc}
\tabletypesize{\scriptsize}
\tablecaption{Metallicity Measurements}
\tablehead{
\colhead{Sightline} & Atomic  & [X/HI]         & [X/H]/[X/H]$_{\odot}$ \\
		    & Species & (uncorrected)  & (corrected)  }
\startdata
Q104816 & \ion{Si}{2}  & -5.1 $\pm$ 0.5 & $ \left. -1.0^{+0.9}_{-0.4} \right.$ \\
Q104816 & \ion{C}{2}   & -4.8 $\pm$ 0.1 & $ \left. -1.6^{+0.5}_{-0.1} \right.$ \\ 
Q104816 & \ion{N}{1}   & -5.2 $\pm$ 0.4 & $ \left. -0.8^{+0.4}_{-0.6} \right.$ \\
Q104709 & \ion{Si}{2}  & -5.1 $\pm$ 0.2 & $ \left. -1.1^{+0.7}_{-0.2} \right.$ \\
Q104709 & \ion{C}{2}   & -4.6 $\pm$ 0.1 & $ \left. -1.0^{+0.7}_{-0.1} \right.$ \\
\enddata
\label{tab:metals} 
\end{deluxetable*}

\clearpage 

\bibliographystyle{astron}


\end{document}